\documentclass[12pt]{iopart}
\def\ket#1{| #1 \rangle}
\def\bra#1{\langle #1 |}

\newtheorem{theorem}{Theorem} 
\newtheorem{lemma}{Lemma}   
\newtheorem{corollary}{Corollary}
\newtheorem{example}{Example}
\newtheorem{definition}{Definition}
\begin{document}
\title{Complete controllability of finite-level quantum systems}
\author{H.\ Fu, S.~G.\ Schirmer and A.~I.\ Solomon}
\address{Quantum Processes Group, The Open University, Milton Keynes, MK7 6AA, UK}
\date{January 23, 2001}

\begin{abstract}
Complete controllability is a fundamental issue in the field of control of quantum 
systems, not least because of its implications for dynamical realizability of the 
kinematical bounds on the optimization of observables.  In this paper we investigate
the question of complete controllability for finite-level quantum systems subject to
a single control field, for which the interaction is of dipole form.  Sufficient 
criteria for complete controllability of a wide range of finite-level quantum systems
are established and the question of limits of complete controllability is addressed.
Finally, the results are applied to give a classification of complete controllability
for four-level systems. 
\end{abstract}
\maketitle

\section{Introduction}
\label{sec:intro}

Recent advances in laser technology have opened up new possibilities for laser control 
of quantum phenomena such as control of molecular quantum states, chemical reaction 
dynamics or quantum computers.  The limited success of initially advocated control 
schemes based largely on physical intuition in both theory and experiment has prompted
researchers in recent years to study these systems using control theory \cite{00RVMK}.

In \cite{98GSLK} it was shown that the kinematical constraint of unitary evolution for
non-dissipative quantum systems gives rise to universal, kinematical bounds on the 
optimization of observables.  It has also been demonstrated that the theoretically 
and practically important question of the dynamical realizability of these universal
bounds depends on the complete controllability of the system \cite{00SL}.  Although 
the issue of complete controllability of quantum systems has been investigated before 
\cite{95RSDRP,00SFS} many open questions remain.

In this paper we study the question of complete controllability of finite-level quantum 
systems driven by a single control, for which the interaction with the control field is
determined by the dipole approximation.  For this kind of system the total Hamiltonian 
is of the form
\begin{equation} \label{eq:H}
  H = H_0 + f(t) H_1,
\end{equation}
where $H_0$ is the internal system Hamiltonian and $H_1$ is the interaction Hamiltonian.
For a finite-level quantum system there always exists a complete orthonormal set of 
energy eigenstates $\ket{n}$ such that $H_0\ket{n}=E_n\ket{n}$ and thus we have
\begin{equation} \label{eq:Hzero}
   H_0 = \sum_{n=1}^N E_n \ket{n}\bra{n} = \sum_{n=1}^N E_n e_{nn}
\end{equation}
where $e_{mn}\equiv \ket{m}\bra{n}$ is an $N\times N$ matrix with elements $(e_{mn})_{kl}
=\delta_{mk}\delta_{nl}$ and $E_n$ are the energy levels of the system.  The $E_n$ are 
real and hence $H_0$ is Hermitian.  The system is non-degenerate provided that $E_n=E_m$
if and only if $m=n$.  The energy levels can be ordered in a non-increasing sequence, 
i.e., $E_1 \le E_2 \le \ldots \le E_N$.  Hence, the energy level spacing 
\begin{equation}
  \mu_n \equiv E_{n+1}-E_n \ge 0, \quad n=1,\ldots,N-1. 
\end{equation}
If $\mu_n=\mu$ for $1\le n \le N-1$ then we say the energy levels are equally spaced. 

Expanding the interaction Hamiltonian $H_1$ with respect to this complete set of 
orthonormal energy eigenstates $\ket{n}$ leads to
\[
   H_1 = \sum_{m,n=1}^N d_{m,n} \ket{m}\bra{n},
\]
where $d_{m,n}$ are the transition dipole moments, which satisfy $d_{m,n}=d_{n,m}^*$,
where $d_{n,m}^*$ is the complex conjugate of $d_{n,m}$.  Thus, $H_1$ is Hermitian.
In the dipole approximation it is generally assumed that only the terms $d_{n-1,n}$ 
and $d_{n,n-1}$ corresponding to transitions between adjacent energy levels are relevant,
i.e., $d_{m,n}=0$ unless $m=n\pm 1$.  Thus, letting $d_n=d_{n,n+1}$ for $1\le n \le N-1$ 
we have
\begin{equation}\label{eq:Hone}
   H_1 = \sum_{n=1}^{N-1} d_n (\ket{n}\bra{n+1}+\ket{n+1}\bra{n})
       = \sum_{n=1}^{N-1} d_n (e_{n,n+1}+e_{n+1,n})
\end{equation}
If the any of the $d_n$ for $1\le n \le N-1$ vanish then the system is decomposable,
i.e., its dynamics can be decomposed into independent subspace dynamics, and therefore 
not completely controllable \cite{00SL}.  Hence, we shall assume that
\begin{equation}
  d_n\neq 0, \quad 1\le n\le N-1, \quad d_0=d_N=0.
\end{equation}
Note that we have introduced the non-physical $d_0=d_N=0$ for convenience.

\section{Sufficient conditions for complete controllability}
\label{sec:lemmas}

\begin{definition}
A quantum system $H=H_0+f(t)H_1$ is \emph{completely controllable} if every unitary 
operator is dynamically accessible from the identity $I$ in $U(N)$ via a path $\gamma(t)
=U(t,t_0)$ that satisfies the equation of motion 
\begin{equation}
  i\hbar \frac{\partial}{\partial t} U(t,t_0)= (H_0+f(t)H_1) U(t,t_0)
\end{equation} 
with initial condition $U(t_0,t_0)=I$.
\end{definition}
In \cite{95RSDRP,72JS} it was shown that a necessary and sufficient condition for 
complete controllability of the system $H=H_0+f(t)H_1$ is that the Lie algebra ${\cal L}$
generated by the skew-Hermitian matrices $\rmi H_0$ and $\rmi H_1$ is $u(N)$, i.e., the 
Lie algebra of skew-hermitian $N\times N$ matrices.  Note that we have $u(N)=su(N)\oplus 
u(1)$ where $su(N)$ is the Lie algebra of traceless skew-Hermitian matrices.  A standard 
basis for $su(N)$ is \cite{62NJ}
\begin{equation}\begin{array}{rcl}
 x_{nn'}&\equiv& e_{nn'}-e_{n'n}, \\
 y_{nn'}&\equiv& \rmi (e_{nn'}+e_{n'n}),\\
 h_n    &\equiv& \rmi (e_{nn}-e_{n+1,n+1}), 
\end{array}\end{equation}
where $1\le n\le N-1$, $n<n'\le N$ and $\rmi=\sqrt{-1}$.  However, to show that ${\cal L}$
contains $su(N)$, it is sufficient to prove that $x_{n,n+1}, y_{n,n+1} \in {\cal L}$ for
$1\le n\le N-1$ since all other basis elements can be generated recursively from 
$x_{n,n+1}$ and $y_{n,n+1}$ for $k>1$:
\begin{eqnarray*}
        x_{n,n+k}&=&[x_{n,n+k-1}, x_{n+k-1,n+k}], \\
        y_{n,n+k}&=&[y_{n,n+k-1}, x_{n+k-1,n+k}], \\ 
        h_n      &=&[x_{n,n+1}, y_{n,n+1}].
\end{eqnarray*}
If $\Tr(H_0)=0$ then the Lie algebra ${\cal L}$ can be at most $su(N)$ since $H_1$ is
traceless by definition.  Note that it can be shown that ${\cal L}=su(N)$ is sufficient
for controllability for many practical purposes \cite{00SFS}.  On the other hand, if
$su(N) \subset {\cal L}$ and $\Tr(H_0)\neq 0$ then $\rmi I$ can be obtained from the
diagonal element $\rmi H_0\in {\cal L}$ since we can write
\begin{equation}
     \rmi H_0= \left[ N^{-1} \mbox{Tr}(H_0) \right] \rmi I + H_0',
\end{equation}
where the traceless matrix $H_0'$ must be a real linear combination of $h_n$ and hence
in the Lie algebra ${\cal L}$.  Thus, if $su(N) \subset {\cal L}$ and $\Tr(H_0)\neq0$
then ${\cal L}=u(N)$.

For a system $H=H_0+f(t)H_1$ with interaction Hamiltonian $H_1$ of the form 
(\ref{eq:Hone}), it turns out that it actually suffices to show that $x_{p,p+1},y_{p,p+1}
 \in {\cal L}$ for some $p$ and $d_{p-k}\neq \pm d_{p+k}$ for some $k$ in order to
conclude that ${\cal L}$ contains $su(N)$.

\begin{lemma} \label{lemma:one}
If $x_{12},y_{12} \in {\cal L}$ then $x_{n,n+1}, y_{n,n+1}\in {\cal L}$ for $1\leq n 
\leq N-1$.  Similarly, if $x_{N-1,N},y_{N-1,N} \in {\cal L}$ then $x_{n,n+1}, y_{n,n+1}
\in {\cal L}$ for $1\leq n \leq N-1$.  
\end{lemma}

\noindent {\bf Proof:} 
Given $x_{12},y_{12}\in {\cal L}$, let $V=\rmi H_1$ and 
\begin{eqnarray*}
 h_1     &\equiv& 2^{-1}[x_{12},y_{12}] =\rmi (e_{11}-e_{22}) \\
 V^{(1)} &\equiv& V-d_1 y_{12}   = \sum_{n=2}^{N-1} d_n y_{n,n+1}.
\end{eqnarray*}  
Since $d_n\neq 0$ for $1\le n \le N-1$ we find that
\[
  d_2^{-1} [h_1, V^{(1)}]=x_{23}\in {\cal L}, \quad [x_{23},h_1]=y_{23}\in {\cal L}. 
\]
By repeating this procedure $N-2$ times, we can show that $x_{n,n+1},y_{n,n+1} \in 
{\cal L}$ for $1\le n\le N-1$.  Similarly, we can prove that given $x_{N-1,N},y_{N-1,N}
\in {\cal L}$ then all $x_{n,n+1}, y_{n,n+1}\in {\cal L}$ for $1\le n \le N-1$. 

\begin{lemma} \label{lemma:two}
If there exists $p$ with $2\le p\le N-2$ such that $x_{p,p+1}, y_{p,p+1} \in {\cal L}$ 
and $k$ such that $d_{p-k} \neq \pm d_{p+k}$ then $x_{n,n+1}, y_{n,n+1} \in {\cal L}$ 
for $1\le n \le N-1$.
\end{lemma}

\noindent {\bf Proof.} 
Given $x_{p,p+1}, y_{p,p+1} \in {\cal L}$ with $2 \le p \le N-2$ then
\[
  h_p\equiv 2^{-1}[x_{p,p+1},y_{p,p+1}] = \rmi (e_{pp} - e_{p+1,p+1}) \in {\cal L}.
\]
Next let $V=iH_1$ and evaluate
\begin{eqnarray*}
 V_p^{(1)}  &\equiv& V-d_p y_{p,p+1}            = \sum_{n\neq p} d_n y_{n,n+1},\\
 X_p^{(1)}  &\equiv& d_{p-1}^{-1}[h_p,V_p^{(1)}]= x_{p-1,p}+\eta_p^{(1)} x_{p+1,p+2},\\
 Y_p^{(1)}  &\equiv& [X_p^{(1)},h_p]            = y_{p-1,p}+\eta_p^{(1)} y_{p+1,p+2},\\
 H_p^{(1)}  &\equiv& 2^{-1}[X_p^{(1)},Y_p^{(1)}]= h_{p-1}  +(\eta_p^{(1)})^2 h_{p+1},\\
{X_p^{(1)}}'&\equiv& 2^{-1}[Y_p^{(1)},H_p^{(1)}]= x_{p-1,p}+(\eta_p^{(1)})^3 x_{p+1,p+2},\\
{Y_p^{(1)}}'&\equiv& 2^{-1}[H_p^{(1)},X_p^{(1)}]= y_{p-1,p}+(\eta_p^{(1)})^3 y_{p+1,p+2},
\end{eqnarray*}  
where $\eta_p^{(1)}=d_{p+1}/d_{p-1}$.  Note that $\eta_p^{(1)}$ is defined and non-zero 
since by hypothesis $d_n\neq 0$ for $1\le n \le N-1$.  This leads to
\begin{eqnarray*}
 (\eta_p^{(1)})^2 X_p^{(1)}-{X_p^{(1)}}' &=& [(\eta_p^{(1)})^2-1] x_{p-1,p}\in {\cal L},\\
 (\eta_p^{(1)})^2 Y_p^{(1)}-{Y_p^{(1)}}' &=& [(\eta_p^{(1)})^2-1] y_{p-1,p}\in {\cal L}.
\end{eqnarray*}
At this point we have to distinguish two cases.  

\noindent {\bf Case 1.}  
If $\eta_p^{(1)}\neq\pm 1$, i.e., $d_{p-1}\neq\pm d_{p+1}$ then it is easy to see that 
$x_{p-1,p},y_{p-1,p}\in {\cal L}$ and hence $h_{p-1}\equiv 2^{-1}[x_{p-1,p},y_{p-1,p}]
\in {\cal L}$ as well.  Now we can proceed to show that $x_{p-2,p-1}, y_{p-2,p-1} \in 
{\cal L}$:
\begin{eqnarray*}
 V_p^{(2)}&\equiv& V_p^{(1)}-d_{p-1} y_{p-1,p}, \\
 X_p^{(2)}&\equiv& d_{p-2}^{-1}[h_{p-1},V_p^{(2)}] =x_{p-2,p-1}\in {\cal L},\\
 Y_p^{(2)}&\equiv& [X_p^{(2)},h_{p-1}]       =y_{p-2,p-1}\in {\cal L}.
\end{eqnarray*}
Repeating the last step $p-2$ times shows that $X_p^{(p-1)}=x_{12}\in {\cal L}$ and 
$Y_p^{(p-1)}=y_{12}\in {\cal L}$ and hence $x_{n,n+1},y_{n,n+1}\in {\cal L}$ for 
$1\le n \le N-1$ by lemma \ref{lemma:one}.

\noindent {\bf Case 2.} 
If $\eta_p^{(1)}=\pm 1$, i.e., $d_{p-1}=\pm d_{p+1}$, then we only have 
${X_p^{(1)}}'=x_{p-1,p}\pm x_{p+1,p+2} \in {\cal L}$ and 
${Y_p^{(1)}}'=y_{p-1,p}\pm y_{p+1,p+2} \in {\cal L}$.  However,
we can now use a similar procedure as above to obtain
\begin{eqnarray*}
 V_p^{(2)}  &\equiv& V_p^{(1)}-d_{p-1}{Y_p^{(1)}}'    
                     = \sum_{n\neq p,p\pm 1} d_n y_{n,n+1},\\ 
 H_p^{(2)}  &\equiv& 2^{-1}[{X_p^{(1)}}',{Y_p^{(1)}}']
                     = h_{p-1}+h_{p+1},\\
 X_p^{(2)}  &\equiv& d_{p-2}^{-1}[H_p^{(2)},V_p^{(2)}]
                     = x_{p-2,p-1}+\eta_p^{(2)}x_{p+2,p+3},\\
 Y_p^{(2)}  &\equiv& [X_p^{(2)},H_p^{(2)}]          
                     = y_{p-2,p-1}+\eta_p^{(2)}y_{p+2,p+3},\\
{H_p^{(2)}}'&\equiv& 2^{-1}[X_p^{(2)},Y_p^{(2)}]   
                     = h_{p-2}+(\eta_p^{(2)})^2 h_{p+2},\\
{X_p^{(2)}}'&\equiv& 2^{-1}[Y_p^{(2)},{H_p^{(2)}}']
                     = x_{p-2,p-1}+(\eta_p^{(2)})^3 x_{p+2,p+3},\\
{Y_p^{(2)}}'&\equiv& 2^{-1}[{H_p^{(2)}}',{X_p^{(2)}}']     
                     = y_{p-2,p-1}+(\eta_p^{(2)})^3 y_{p+2,p+3},
\end{eqnarray*}
where $\eta_p^{(2)}=d_{p+2}/d_{p-2}$.  This leads to
\begin{eqnarray*}
 (\eta_p^{(2)})^2 X_p^{(2)}-{X_p^{(2)}}'&=& [(\eta_p^{(2)})^2-1] x_{p-2,p-1}\in {\cal L},\\
 (\eta_p^{(2)})^2 Y_p^{(2)}-{Y_p^{(2)}}'&=& [(\eta_p^{(2)})^2-1] y_{p-2,p-1}\in {\cal L}.
\end{eqnarray*}
Again, we have to consider two different cases.  

\noindent {\bf Case 2a:} If $\eta_p^{(2)}\neq \pm 1$, i.e., $d_{p-2}\neq \pm d_{p+2}$, 
then $x_{p-2,p-1},y_{p-2,p-1}\in {\cal L}$ as well as 
$ h_{p-2}\equiv 2^{-1}[x_{p-2,p-1}, y_{p-2,p-1}] \in {\cal L}$ and we can proceed as in 
case 1 to show that $x_{p-3,p-2},y_{p-3,p-2} \in {\cal L}$:
\begin{eqnarray*}
 V_p^{(3)} &\equiv& V_p^{(2)}-d_{p-2} y_{p-2,p-1}, \\
 X_p^{(3)} &\equiv& d_{p-3}^{-1}[h_{p-2},V_p^{(3)}] = x_{p-3,p-2}\in {\cal L},\\
 Y_p^{(3)} &\equiv& [X_p^{(3)},h_{p-2}]       = y_{p-3,p-2}\in {\cal L}.
\end{eqnarray*}
Repeating the last step $p-3$ times shows that $X_p^{(p-1)}=x_{12} \in {\cal L}$
and $Y_p^{(p-1)}=y_{12}\in {\cal L}$ and hence $x_{n,n+1},y_{n,n+1} \in {\cal L}$ 
for $1 \le n \le N-1$ by lemma \ref{lemma:one}.

\noindent {\bf Case 2b:} If $\eta_p^{(2)}=\pm1$, i.e., $d_{p-2}=\pm d_{p+2}$, then we 
have only $X_p^{(2)}=x_{p-2,p-1}\pm x_{p+2,p+3}\in {\cal L}$ and $Y_p^{(2)}= y_{p-2,p-1}
\pm y_{p+2,p+3}\in {\cal L}$ but we can proceed as in case 2 to obtain
\begin{eqnarray*}
 {X_p^{(3)}}' &=& x_{p-3,p-2}+(\eta_p^{(3)})^3 x_{p+3,p+4}\in {\cal L},\\
 {Y_p^{(3)}}' &=& y_{p-3,p-2}+(\eta_p^{(3)})^3 y_{p+3,p+4}\in {\cal L},
\end{eqnarray*}
where $\eta_p^{(3)}=d_{p+3}/d_{p-3}$.  Again, we must distinguish the cases 
$\eta_p^{(3)}\neq \pm 1$ and $\eta_p^{(3)}=\pm 1$ and so forth.

Using this procedure, we can always show that $x_{p-k,p-k+1},y_{p-k,p-k+1} \in {\cal L}$
since by hypothesis $d_{p-k}\neq \pm d_{p+k}$.  We can then proceed as in case 1 to show 
that $x_{12},y_{12} \in {\cal L}$, from which it follows that all $x_{n,n+1}, 
y_{n+1,n} \in {\cal L}$ by lemma 1. QED.

\section{Completely controllable quantum systems}
\label{sec:controllability}

\subsection{Anharmonic Systems}

The results of the previous section can be applied to establish complete controllability
for many quantum systems.

\begin{theorem}\label{thm:one}
The dynamical Lie algebra for a quantum system $H=H_0+f(t)H_1$ with $H_0$ and $H_1$ as 
in \eref{eq:Hzero} and \eref{eq:Hone} is at least $su(N)$ if either
\begin{enumerate}
\item $\mu_1\neq 0$ and $\mu_n \neq \mu_1$ for $2\le n\le N-1$, or 
\item $\mu_{N-1}\neq 0$ and  $\mu_n \neq \mu_{N-1}$ for $1\le n \le N-2$.
\end{enumerate}
If in addition $\Tr(H_0)\neq 0$ then the dynamical Lie algebra is $u(N)$, i.e., the
system is completely controllable.
\end{theorem}   

\noindent {\bf Proof:} Suppose $\mu_1\neq 0$ and $\mu_n \neq \mu_1$ for $2\le n\le N-1$.  
Let $V=\rmi H_1$ and evaluate
\begin{eqnarray*} 
 V'       &\equiv& [\rmi H_0, V]   =\sum_{n=1}^{N-1}\mu_n   d_n x_{n,n+1}  \label{vp} \\
 V''      &\equiv& [V',\rmi H_0]   =\sum_{n=1}^{N-1}\mu_n^2 d_n y_{n,n+1}  \label{vpp}\\
 V^{(1)}  &\equiv& V''-\mu_{N-1}^2 V=\sum_{n=1}^{N-2}(\mu_n^2-\mu_{N-1}^2) d_n y_{n,n+1}\\
 V^{(2)}  &\equiv& [[\rmi H_0, V^{(1)}],\rmi H_0]-\mu_{N-2}^2 V^{(1)} \\
          &=& \sum_{n=1}^{N-3}(\mu_n^2-\mu_{N-2}^2)(\mu_n^2-\mu_{N-1}^2) d_n y_{n,n+1} \\
          &\vdots& \\
 V^{(k)}  &\equiv& [[\rmi H_0, V^{(k-1)}],\rmi H_0]-\mu_{N-2}^2 V^{(k-1)} \\
          &=& \sum_{n=1}^{N-1-k} 
              \left[\prod_{k=n+1}^{N-1} d_n (\mu_n^2-\mu_k^2)\right] y_{n,n+1} \\
          &\vdots& \\
 V^{(N-2)}&\equiv& d_1\left[\prod_{k=2}^{N-1}(\mu_1^2-\mu_k^2)\right] y_{12} \in {\cal L}.
\end{eqnarray*}
Since by hypothesis $d_1\neq 0$, $\mu_1\neq 0$ and $\mu_n\neq\mu_1$ for $2\le n\le N-1$ 
we have $y_{12}\in {\cal L}$ and hence $\mu_1^{-1}[\rmi H_0,y_{12}]=x_{12} \in {\cal L}$.
Hence, $su(N) \subset {\cal L}$ by lemma 1 and if $\Tr(H_0)\neq 0$ then ${\cal L}=u(N)$.
The proof for the case $\mu_n \neq \mu_{N-1}$ for $1\le n \le N-2$ is analogous. QED

This theorem, first proved in \cite{00SFS}, is very important in that it guarantees the
complete controllability of physically important systems such as simple atomic systems
or Morse oscillators, which are often used to model molecular bonds.

\begin{example} \label{ex:Morse}
The energy level spacing for a Morse oscillator is of the form $\mu_n\propto 1-B n$ 
where $B$ is a small positive real number and we can assume $\Tr(H_0)\neq 0$. Therefore,
$\mu_n\neq \mu_1$ for $n>1$ and thus any $N$-level Morse oscillator system is completely
controllable. 
\end{example}

\noindent Theorem \ref{thm:one} also applies to degenerate or more complicated systems.
\begin{example} \label{ex:degen}
Consider a system with energy levels $E_1$ and $E_n=E_2 \neq E_1$ for $2 \le n\le N$ 
with arbitrary non-zero transition dipole moments $d_n$.  In this case we have $\mu_1
=E_2-E_1\neq 0$ but $\mu_n=0$ for $2\le n\le N-1$.  Thus, surprisingly, this highly
degenerate system is completely controllable by theorem \ref{thm:one} provided that
$\Tr(H_0)\neq 0$.
\end{example}

\begin{example} \label{ex:atomic}
The energy levels of the bound states of a one-electron atom of atomic number $Z$ are 
$E_n=(-13.9\, {\rm eV}) Z^2/n^2$.  Therefore, $\Tr(H_0)\neq0$ and the energy level 
spacing is 
\[
  \mu_n \propto n^{-2}-(n+1)^{-2} = \frac{(2n+1)}{n^2(n+1)^2}.
\]
Note that the multiplicity of energy level $E_n$ is $2n^2$ including angular momentum
and spin degeneracy, i.e., the energy levels are degenerate.  Nevertheless, we can apply
theorem \ref{thm:one} to conclude that any non-decoupled $N$-level subsystem of this 
model consisting of at least two different energy levels is completely controllable if 
the interaction with the control field is of dipole form (\ref{eq:Hone}).
\end{example}

\begin{example} \label{ex:box}
The energy levels $E_n$ for a particle in a 1D box are $C n^2$, where $C$ is a positive
constant.  Hence, $\Tr(H_0)\neq 0$ and $\mu_n \neq \mu_1$ for $n>1$, i.e., any 
non-decoupled $N$-level subsystem of this model is completely controllable according 
to theorem \ref{thm:one} if the interaction with the control field is of dipole form 
(\ref{eq:Hone}).
\end{example}

\begin{example} \label{ex:sysa}
Consider a $N=2\ell+1$ level system with $\Tr(H_0)\neq 0$ and energy level spacings 
$\mu_{2k}=\mu_2$ for $1\le k\le \ell$ but $\mu_1 \neq \mu_n$ for $n>1$.  This system 
is also completely controllable (independent of the $d_n$) according to theorem 
\ref{thm:one}.  Similarly, if we had $\mu_{2k-1}=\mu_1$ for $1\le k\le \ell$ but 
$\mu_{2\ell}\neq\mu_n$ for $n<2\ell$ then the system would be completely controllable 
according to theorem \ref{thm:one} as well.
\end{example}

The last example is interesting for the following reason.  Suppose we considered 
instead a composite system of $\ell$ coupled identical two-level systems with simple 
interactions, i.e., an $N=2\ell$ level system with energy level spacings $\mu_{2k+1}
=\mu_1$ for $1<k<\ell$ but, e.g., $\mu_2\neq\mu_n$ for $n\neq 2$.  In this case we 
have $\mu_1=\mu_{2\ell-1}$, i.e., theorem \ref{thm:one} does not apply although 
considering the last example one would expect this system to be controllable as well.
This suggests that theorem \ref{thm:one} can be generalized.

\begin{theorem}  \label{thm:two}
The dynamical Lie algebra ${\cal L}$ of a quantum system $H=H_0+f(t)H_1$ with $H_0$ and
$H_1$ as in \eref{eq:Hzero} and \eref{eq:Hone} is at least $su(N)$ if there exists 
$\mu_p\neq 0$ such that $\mu_n\neq \mu_p$ for $n\neq p$, and $k$ such that $d_{p-k}
\neq\pm d_{p+k}$.  If in addition $\Tr(H_0)\neq 0$ then ${\cal L}=u(N)$, i.e., the
system is completely controllable.
\end{theorem}

\noindent {\bf Proof:} Let $V=iH_1$ and define
\begin{eqnarray*}
 V^{(1)}  &\equiv& [[\rmi H_0, V],\rmi H_0]-\mu_{\sigma(1)}^2 V \\
 V^{(2)}  &\equiv& [[\rmi H_0, V^{(1)}],\rmi H_0]-\mu_{\sigma(2)}^2 V^{(1)} \\
 \vdots \\
 V^{(N-2)}&\equiv& [[\rmi H_0, V^{(N-3)}],\rmi H_0]-\mu_{\sigma(N-3)}^2 V^{(N-3)} 
\end{eqnarray*}
where $\sigma$ is a permutation of the set $\{1,2,\ldots,N-1\}$ such that
$\sigma(N-1)=p$.  Then 
\[
 V^{(N-2)}= d_p \left[\prod_{n=1}^{N-2}(\mu_p^2-\mu_{\sigma(n)}^2)\right] y_{p,p+1} 
\in {\cal L}.
\] 
and since by hypothesis $d_p\neq 0$, $\mu_p\neq 0$ and $\mu_n\neq\mu_p$ for $n\neq p$ 
we have $y_{p,p+1}\in {\cal L}$ and hence $\mu_p^{-1}[\rmi H_0,y_{p,p+1}]=x_{p,p+1}\in 
{\cal L}$.  By hypothesis we have furthermore $d_{p+k}\neq\pm d_{p-k}$ for some $k$.
Hence, $su(N) \subset {\cal L}$ by lemma 2 and if $\Tr(H_0)\neq 0$ then ${\cal L}=u(N)$,
i.e., the system is completely controllable. QED

Note that $d_0=d_N=0$ and $d_n\neq 0$ for $1\le n\le N-1$ implies that the condition
$d_{p+k}\neq \pm d_{p-k}$ is always satisfied for $k=\min\{p,N-p\}$ unless $p=N-p$ 
since if $k=p<N-p$ then $d_{p-k}=d_0=0$ and $d_{p+k}\neq 0$, and if $k=N-p<p$ then 
$d_{p+k}=d_N=0$ and $d_{p-k}\neq 0$.  Furthermore, $p=N-p$ is only possible if $p=N/2$
and hence $N$ even.  Thus, assuming $\Tr(H_0)\neq 0$, we have the following 

\begin{corollary} \label{cor:one}
If $N$ is odd and there exists $\mu_p\neq 0$ such that $\mu_n\neq\mu_p$ for $n\neq p$, 
then the system is completely controllable. 
\end{corollary}

\begin{corollary} \label{cor:two}
If $N$ is even and there exists $\mu_p\neq 0$ with $p\neq N/2$ such that $\mu_n\neq 
\mu_p$ for $n\neq p$, then the system is completely controllable.
\end{corollary}

Applying corollary \ref{cor:two} to the $N=2\ell$ level composite system with energy 
level spacings $\mu_{2k+1}=\mu_1$ for $1<k<\ell$ but $\mu_2 \neq \mu_n$ for $n\neq 2$
considered above, we see that the system is always controllable for $N\neq 4$.  
If $N=4$ then it is controllable if $d_1 \neq \pm d_3$.  There are many other 
applications for the theorems and corollaries above.

\begin{example} \label{ex:two}
The system of two coupled $\ell$-level harmonic oscillators with
\begin{equation} \label{eq:En}
 E_n=\cases{E_1+(n-1)\mu        & for $1\le n\le \ell$  \\
            E_1+(n-1)\mu+\Delta & for $\ell+1\le n\le 2\ell$\\} 
\end{equation}
is completely controllable if $d_{\ell-k}\neq \pm d_{\ell+k}$ for some $k$ since 
$\mu_n=\mu$ for $n\neq \ell$ and $\mu_\ell=\mu+\Delta$.  For instance, if
\begin{equation} \label{eq:d1}
  d_n=\cases{\sqrt{n}      & for $1\le n\le\ell-1$ \\
             d\neq 0     & for $n=\ell$ \\
             \sqrt{n-\ell} & for $\ell+1\le n\le 2\ell-1$\\}
\end{equation}
then the system is completely controllable by theorem \ref{thm:two} since, e.g., 
$d_{\ell-1}=\sqrt{\ell-1}\neq \sqrt{1}=d_{\ell+1}$.  However, if 
\begin{equation} \label{eq:d2}
  d_n=\cases{1       & for $1\le n\le\ell-1$ \\
             d\neq 0 & for $n=\ell$ \\
             1       & for $\ell+1\le n\le 2\ell-1$\\}
\end{equation}
then the system does not satisfy the criteria for complete controllability established
in the previous theorems and one can verify that the system is indeed not completely
controllable.
\end{example}

\subsection{Harmonic oscillators}

Theorem \ref{thm:two} and its corollaries establish complete controllability for many 
anharmonic, non-decomposable quantum systems.  However, the conditions on the $\mu_n$ 
exclude systems with equally spaced energy levels, i.e., $\mu_n=\mu$ for $1\le n\le N-1$,
such as harmonic oscillators.  For these systems we can not apply the techniques used 
in the previous section to establish complete controllability since in this case 
$[[\rmi H_0, V],\rmi H_0]=\mu V$.  To resolve this problem, we introduce a new set of 
parameters depending on the values of the transition dipole moments $d_n$
\begin{equation} \label{vn}
     v_n= 2d_n^2-d_{n-1}^2-d_{n+1}^2, \quad 1\le n \le N-1,
\end{equation}
which determine whether the system is completely controllable or not.

\begin{theorem}  \label{thm:three}
The dynamical Lie algebra ${\cal L}$ for a system $H=H_0+f(t)H_1$ with equally spaced 
energy levels, i.e., $\mu_n=\mu\neq 0$ for $1\le n\le N-1$ is at least $su(N)$ if there
exists $v_p\neq 0$ such that $v_n \neq v_p$ for $n\neq p$ and $p\neq N/2$; if $p=N/2$ 
then $d_{p-k}\neq \pm d_{p+k}$ for some $k$ is required as well.  If in addition 
$\Tr(H_0)\neq 0$ then ${\cal L}=u(N)$.
\end{theorem}

\noindent{\bf Proof: }  For convenience we define $Y^{(1)} \equiv iH_1$. Then we have 
\begin{eqnarray*}
 X^{(1)} &\equiv& \mu^{-1}[iH_0,Y^{(1)}]   =     \sum_{n=1}^{N-1} d_n x_{n,n+1}, \\
 Z       &\equiv& 2^{-1}  [X^{(1)},Y^{(1)}]= \rmi\sum_{n=1}^N (d_n^2-d_{n-1}^2) e_{nn}.
\end{eqnarray*}
From $X^{(1)}, Y^{(1)}$ and $Z^{(1)}$, we have
\begin{eqnarray*}
 Y^{(2)} &\equiv& [Z, X^{(1)}]-v_{\sigma(1)} Y^{(1)}
                = \sum_{n=2}^{N-1} (v_n-v_{\sigma(1)}) d_n y_{n,n+1}\\
 X^{(2)} &\equiv& [Y^{(1)}, Z]-v_{\sigma(1)} X^{(1)}
              = \sum_{n=2}^{N-1} (v_n-v_{\sigma(1)}) d_n x_{n,n+1}\\
 Y^{(3)} &\equiv& [Z, X^{(2)}]-v_{\sigma(2)} Y^{(2)}
              = \sum_{n=3}^{N-1} (v_n-v_{\sigma(1)})(v_n-v_{\sigma(2)}) d_n y_{n,n+1} \\
 X^{(3)} &\equiv& [Y^{(2)}, Z]-v_{\sigma(2)} X^{(2)}
              = \sum_{n=3}^{N-1} (v_n-v_{\sigma(1)})(v_n-v_{\sigma(2)}) d_n x_{n,n+1} \\
 \vdots \\
 Y^{(N-1)} &\equiv& [Z,X^{(N-2)}]-v_{\sigma(N-2)} Y^{(N-2)}
        = d_p\left[ \prod_{n=1}^{N-2} (v_p-v_{\sigma(n)})\right] y_{p,p+1} \in {\cal L},\\
 X^{(N-1)} &\equiv& [Y^{(N-2)},Z]-v_{\sigma(N-2)} X^{(N-2)}
        = d_p\left[ \prod_{n=1}^{N-2} (v_p-v_{\sigma(n)})\right] x_{p,p+1} \in {\cal L}.
\end{eqnarray*}
where $\sigma$ is a permutation of the set $\{1,2,\ldots,N-1\}$ such that $\sigma(N-1)=p$.
By hypothesis we have $v_p \neq 0$ and $v_n\neq v_p$ for $n\neq p$.  Hence, we can 
conclude $x_{p,p+1}, y_{p,p+1} \in {\cal L}$.  If $p\neq N/2$ then the condition 
$d_{p-k}\neq \pm d_{p+k}$ is automatically satisfied for $k=\mbox{min}\{p,N-p\}$; 
otherwise it is guaranteed by the hypothesis of the theorem.  Therefore, ${\cal L}$
contains $su(N)$ by lemma \ref{lemma:two} and if $\Tr(H_0)\neq 0$ then ${\cal L}=u(N)$.
QED

\begin{example} \label{ex:harmonic}
The truncated $N$-level harmonic oscillator with $E_n\propto n+\frac{1}{2}$ and 
$d_n=\sqrt{n}$ is completely controllable since $v_n=0$ for $1\le n\le N-2$ but 
$v_{N-1}=N\neq 0$.
\end{example}

\begin{example} \label{ex:harmonic2}
A system with $N$ equally spaced energy levels, $\Tr(H_0)\neq 0$, and $d_n=1$ for 
$1\le n\le N-2$ and $d_{N-1}\neq \pm 1$ is completely controllable since $v_1=1$, 
$v_n=0$ for $2\le n\le N-3$, $v_{N-2}=1-d_{N-1}^2$ and $v_{N-1}=2d_{N-1}^2-1\neq 0$, 
i.e., $v_1 \neq v_n$ for $n>1$.
\end{example}

\section{Limits of complete controllability}
\label{sec:limits}

The theorems and corollaries in section \ref{sec:controllability} suggest that many 
non-decomposable quantum systems are completely controllable and one might actually 
begin to wonder if there are any non-decomposable (i.e., non-decoupled) systems that 
are not completely controllable.  Unfortunately, the answer is yes, and worse yet, 
these systems may look very similar to completely controllable systems.  Recall 
example \ref{ex:two}, i.e., a system with energy levels (\ref{eq:En}), which satisfies
the conditions for complete controllability of theorem \ref{thm:two} if the transition
dipole moments are chosen as in (\ref{eq:d1}) but does not satisfy the criteria for 
complete controllability if the $d_n$ are chosen as in (\ref{eq:d2}).  We shall see 
that for the latter choice of the transition dipole moments $d_n$ the system is indeed
\emph{not} completely controllable.

\begin{theorem}\label{thm:four}
The dynamical Lie algebra for a system with $N$ equally spaced energy levels $\mu_n=\mu$
for $1\le n\le N-1$ and $v_n=v$ for $1\le n\le N-1$ has dimension four, i.e., the system
is not completely controllable for $N>2$.
\end{theorem}     

\noindent {\bf Proof:} Let $Y=iH_1$,
\begin{eqnarray*} 
 X &\equiv& \mu^{-1}[\rmi H_0,Y]=     \sum_{n=1}^{N-1}  d_n x_{n,n+1},\\
 Z &\equiv& 2^{-1} [X,Y]        = \rmi\sum_{n=1}^N (d_n^2-d_{n-1}^2) e_{nn}
\end{eqnarray*}
and note that we have the following commutation relations
\begin{eqnarray*}
 [\rmi H_0, X]=\mu Y, &\quad [\rmi H_0, Y]=-\mu X, \quad & [\rmi H_0, Z]=0, \\
 {[X, Y]=2Z,}         &\quad [Z,X]=-v_1 Y,         \quad & [Z,Y]=v_1 X.
\end{eqnarray*} 
Hence, $\rmi H_0,X,Y,Z$ span the Lie algebra ${\cal L}$.  Thus ${\cal L}$ is 
isomorphic to the 4-dimensional Lie algebra $u(2)$, i.e., the system is not completely 
controllable for $N>2$. QED

\begin{example} \label{ex:N3}
If $N=3$ and $\mu_1=\mu_2$ as well as $d_1=d_2$ then the system is not completely 
controllable since $v_1=2d_1^2-d_2^2=2d_2^2-d_1^2=v_2$, i.e., the Lie algebra has 
dimension four according to the previous theorem.
\end{example}

Theorem \ref{thm:four} has another important implication.  One can prove by induction 
that for $N>4$ the condition $v_n=v$ for $1\le n\le N-1$ can only be satisfied if
\[
  d_n^2 = n d_1^2-\frac{n(n-1)}{2} v, \quad {\rm and} \quad
  v= \frac{2}{N-4} d_1^2.
\]
Note that $v\rightarrow 0$ and thus $d_n\rightarrow n d_1^2$ for $N\rightarrow\infty$.
Hence, the infinite dimensional harmonic oscillator with $d_n=\sqrt{n}$ for $1\le n\le 
\infty$ is \emph{not} completely controllable.

\begin{theorem} \label{thm:five}
A system with $N$ equally spaced energy levels and $d_n=1$ for $1\le n\le N-1$ is 
not completely controllable for $N>2$.
\end{theorem}

\noindent {\bf Proof}: Note that theorem \ref{thm:three} does not apply since 
$v_1=v_{N-1}=1$ and $v_2=\cdots =v_{N-2}=0$.  Setting
\begin{eqnarray*}
     Z_1 &=& i (e_{11} - e_{nn}), \\
     X_1 &=& \mu^{-1}[iH_0,iH_1]= \sum_{n=1}^{N-1} x_{n,n+1}
\end{eqnarray*}
leads to $[Z_1,X_1]=y_{12}+y_{N-1,N}\in {\cal L}$.  Thus, we have the following 
reduced problem 
\[
     \rmi H_0, \qquad
     \rmi H_1^{(1)}= \rmi H_1-[Z_1,X_1]= \sum_{n=2}^{N-2} y_{n,n+1}.
\]
and we can use induction to prove the theorem.  One can easily verify that the system 
is not completely controllable if $N=3$ (see example \ref{ex:N3}) or $N=4$ (see section
\ref{sub:caseA}).  Suppose the theorem is true for $N>2$.  We want to prove it is also 
true for $N+2$.  To this end, assume that the theorem is not true, i.e., that the 
$N+2$ level system is completely controllable.  Then we can use the above produre to 
reduce the system to an $N$-level system.  Clearly, this reduced $N$ level system must 
be completely controllable, which contradicts the assumption.  This means that the 
assumption is false. QED 

\section{Classification of complete controllability for four-level systems}
\label{sec:classification}

In this section we apply the results of sections \ref{sec:controllability} and
\ref{sec:limits} to give a classification of the complete controllability problem 
for four-level quantum systems whose interaction with the control field is determined 
by (\ref{eq:Hone}).

According to theorem \ref{thm:two} and its corollaries, a non-decomposable four-level
quantum system is completely controllable if $\Tr(H_0)\neq0$ and one of the following 
conditions apply:
\begin{enumerate}
\item  $\mu_1 \neq \mu_2 \neq \mu_3 $;
\item  $\mu_1\neq \mu_2 = \mu_3 $;
\item  $\mu_1 = \mu_2 \neq \mu_3 $;
\item  $\mu_1 = \mu_3\neq \mu_2 $, $\mu_2\neq 0$ and $d_1\neq\pm d_3$,
\end{enumerate}
Furthermore, if $\mu_1=\mu_2=\mu_3\neq 0$ then the system is completely controllable 
according to theorem \ref{thm:three} if either
\begin{enumerate}
\item $v_1\neq v_2, v_1\neq v_3$;
\item $v_3\neq v_1, v_3\neq v_2$. 
\end{enumerate}
Decomposable systems (and traceless systems) are not completely controllable.  Hence, 
the only cases that remain to be considered are the following.

\subsection{Case $\mu_1=\mu_3\neq\mu_2 $, $\mu_2\neq 0$, $d_1=\pm d_3$} 
\label{sub:caseA}
In this case the system is {\em not} completely controllable and we actually have an 
11-dimensional Lie algebra isomorphic to $sp(2)\oplus u(1)$.  To see this, note that 
$sp(2)$ is spanned by \cite{62NJ}
\begin{equation} \label{eq:Cl}
\begin{array}{rclrcl}
  h_1                   &\equiv& \rmi (e_{11}-e_{33}) \qquad & 
  h_2                   &\equiv& \rmi (e_{22}-e_{44}) \\
  x_{2\omega_1}         &\equiv& x_{31}              \qquad & 
  y_{2\omega_1}         &\equiv& y_{31}              \\
  x_{2\omega_2}         &\equiv& x_{42}              \qquad & 
  y_{2\omega_2}         &\equiv& y_{42}              \\
  x_{\omega_1+\omega_2} &\equiv& x_{32}+x_{41}      \qquad &
  y_{\omega_1+\omega_2} &\equiv& y_{32}+y_{41}      \\
  x_{\omega_1-\omega_2} &\equiv& x_{21}-x_{34}      \qquad&
  y_{\omega_1-\omega_2} &\equiv& y_{21}-y_{34}  
\end{array}
\end{equation}
and if $d_1=\pm d_3$ then the change of basis
\[
 \{|1\rangle, |2\rangle, |3\rangle, |4\rangle\} \mapsto 
 \{|2\rangle, |1\rangle, |3\rangle, \mp |4\rangle\}
\]
leads to 
\begin{eqnarray*}
  H_0 &=& \frac{1}{4} \Tr(H_0) I - (\mu_1+\mu_2/2) h_1+ \mu_2 h_2 \\
  H_1 &=& d_1 y_{\omega_1-\omega_2}+ d_2 y_{2\omega_1}
\end{eqnarray*}
and one can easily check that $H_0'=\mu_1 h_1+\mu_2 h_2$ and $H_1$ generate all of 
$sp(2)$.  Hence, ${\cal L}\simeq sp(2)\oplus u(1)$.

\subsection{Case $\mu_1=\mu_3\neq \mu_2$, $\mu_2 = 0$}
\label{sub:caseB}
\begin{eqnarray*}
 X_1 &\equiv& \mu_1^{-1} [\rmi H_0,\rmi H_1] = d_1 x_{12} + d_3 x_{34}, \\
 Y_1 &\equiv& \mu_1^{-1} [\rmi H_0, X_1] = d_1 y_{12} + d_3 y_{34}, \\
 Z_1 &\equiv& 2^{-2}[X_1,Y_1] = \rmi[d_1^2(e_{11}-e_{22})+d_3^2 (e_{33}-e_{44})].        
\end{eqnarray*}
Then we have
\begin{eqnarray*}
 Y_2 &\equiv& d_2^{-1}(\rmi Z_1-Y_1)      = y_{23} \in {\cal L},\\
 X_2 &\equiv& (d_1^2+d_2^2)^{-1}[Z_1,Y_2] = x_{23} \in {\cal L}.
\end{eqnarray*}
According to lemma 2, the system is completely controllable if $d_1\neq\pm d_3$.  For 
$d_1=\pm d_3$, the Lie algebra generated is the 11-dimensional Lie algebra given in the 
previous section, i.e., the system is not completely controllable. 

\subsection{Case $\mu_1=\mu_2=\mu_3$, $v_1=v_3$, $v_2\neq v_1$}
\label{sub:caseC}
From the definition of $v_n$ we obtain that the condition $v_2\neq v_1=v_3$ is 
equivalent to $d_2^2 \neq d_1^2=d_3^2$, which implies $d_1=\pm d_3$.  One can easily
verify that ${\cal L}$ is the 11-dimensional algebra given in section \ref{sub:caseA}.
Hence, the system is not completely controllable.

\subsection{Case $\mu_1=\mu_2=\mu_3$, $v_1=v_2=v_3$}
\label{sub:caseD}
In this case one can easily verify that $\rmi H_0$ and $\rmi H_1$ generate a Lie algebra
of dimension four isomorphic to $u(2)$. Hence, the system is not completely controllable.
Note that $v_1=v_2=v_3$ is equivalent to $d_3^2=d_1^2=\frac{3}{4} d_2^2$.  (See also 
theorem \ref{thm:five})

\subsection{Case $\mu_1=\mu_2=\mu_3=0$}
\label{sub:caseE}
This is a completely degenerate system ($E_1=E_2=E_3=E_4$) and ${\cal L}$ is a 
two-dimensional Lie algebra with basis $\rmi H_0=\rmi I$ and $\rmi H_1$.  Clearly, 
the system is not completely controllable.

\begin{table}
\begin{center}
\begin{tabular}{|c|c|c|c|} \hline
System & $\mu_n$ & $d_n (\neq 0)$ & Complete controllability \\ \hline\hline
       & $\mu_1\neq\mu_2\neq\mu_3$ & & Yes \\ \cline{2 - 4}
       & $\mu_1\neq\mu_2 =   \mu_3$ & & Yes \\ \cline{2 - 4}
AO     & $\mu_1 = \mu_2 \neq \mu_3$ & & Yes \\ \cline{2 - 4}
       & $\mu_1 = \mu_3 \neq \mu_2$ & $d_1\neq d_3$& Yes \\ \cline{3 - 4}
       &                            & $d_1= d_3$& No \\ \hline\hline
HO     & & $v_1\neq v_2 \neq v_3$ & Yes \\ \cline{3 - 4}
       & &$v_1\neq v_2 = v_3$ & Yes \\ \cline{3 - 4}             
       & $\mu_1 = \mu_3 = \mu_3\neq 0 $ & $v_1= v_2 \neq v_3$ & Yes \\ \cline{3 - 4} 
       & & $v_1= v_3 \neq v_2$, $d_1=\pm d_3$& No \\ \cline{3 - 4}
       & &$v_1=v_2 = v_3$ & No \\ \cline{2 - 4}
       &$\mu_1 = \mu_3 = \mu_3=0$ & & No \\ \hline                     
\end{tabular} 
\end{center}
\caption{Complete controllability for four level system}
\label{table:one}
\end{table}

\section{Conclusion}
\label{sec:conclusion}

In this paper we studied the problem of complete controllability of finite-level
non-decomposable quantum systems whose interaction with a semi-classical field is
governed by the dipole approximation.  We reduced the problem of complete controllability
of these systems to the question whether a pair of skew-Hermitian matrices $x_{n\,n+1}$ 
and $y_{n\,n+1}$ can be generated by $\rmi H_0$ and $\rmi H_1$.  Using these criteria, 
we showed that many non-decomposable finite-level quantum systems are completely 
controllable, including many atomic systems as well as Morse and harmonic oscillator 
systems.  We also showed however that complete controllability is by no means a universal
property of the types of systems under consideration and that in fact many systems
lacking complete controllability may appear superficially very similar to completely
controllable ones.  Finally, we applied our results to give a classification of 
four-level systems in terms of complete controllability.

\ack
We are grateful to Prof.\ Zhaoyan Wu and Dr.\ V.~Dobrev for their useful suggestions.
H.~Fu is supported in part by the National Natural Science Foundation of China 
through Northeast Normal University (19875008) and the State Key Laboratory of 
Theoretical and Computational Chemistry, Jilin University.

\end{document}